\documentclass [11pt,a4paper] {article}

\usepackage[cp1252]{inputenc}
\usepackage[english]{babel}

\usepackage{amssymb}
\usepackage{amsmath}
\usepackage{amsfonts,amssymb}

\usepackage[dvips]{graphicx}

\begin{document}

\title{Applying computational complexity\\to the emergence of classicality}

\author{Arkady Bolotin\footnote{$Email: arkadyv@bgu.ac.il$} \\ \textit{Ben-Gurion University of the Negev, Beersheba (Israel)}}

\maketitle

\begin{abstract}\noindent Can the computational complexity theory of computer science and mathematics say something new about unresolved problems in quantum physics? Particularly, can the \textbf{P} versus \textbf{NP} question in the computational complexity theory be a factor in the elucidation of the emergency of classicality in quantum mechanics? The paper compares two different ways of deriving classicality from the quantum formalism resulted from two differing hypotheses regarding the \textbf{P} versus \textbf{NP} question -- the approach of the quantum decoherence theory implying that \textbf{P }= \textbf{NP }and the computational complexity approach which assumes that \textbf{P }is not equal to \textbf{NP}.\\

\end{abstract}

\section{Introduction}

\noindent Let us consider a system of $N$ qubits (i.e. a quantum system comprised of $N$ entangled spin-½ particles). Such a system can be in an arbitrary superposition of up to $2^N$ different states simultaneously; therefore, in general, a quantum state of this system is specified by $2^N$ complex numbers, that is, probability amplitudes of the states (one for every possible outcome of measuring the spins in the \{0, 1\} basis). Hence, even for the modest value of, say, $N=500$ (this may be a system containing just a few hundred atoms) the number of the amplitudes will be larger than the estimated number of atoms in the whole Universe. On the other hand, it is the time-dependent Schrödinger equation that gives the description of the system's quantum state (the vector in a space of $2^{500}$ dimension) evolving with time. So, the question becomes, ``How Nature can manipulate such enormous data that fast -- i.e. in parallel with the system's evolution?'' As Nielsen and Chuang put this, ``It is as if Nature were keeping $2^{500}$ hidden pieces of scratch paper on the side, on which she performs her calculations as the system evolves'' \cite{Nielsen}.\\

\noindent If Nature can do it, so can we. This was the essence of the visionary idea of a quantum computer suggested by Feynman -- to take advantage of the Nature's (allegedly) enormous computational power in order to perform simulations of quantum mechanical systems, extremely difficult to simulate on a classical computer \cite{Feynman}.\\

\noindent But let us for a while leave aside questions of whether a large-scale quantum computer can be ever built or what the reason for the quantum computational speedup is (which are usually being asked in connection with quantum computation) and instead ask this. What if Nature could not manipulate such enormous quantities of data in parallel with the system's evolution (or within any reasonable time) at all? What if her computational power was actually limited in such a way that solving the Schrödinger equation prescribed by a system with a huge number of variables needed to describe any possible state of the system (like $2^{500}$ amplitudes) would be unfeasible even for Nature and thus this equation might be resolved in a short time only approximately?\\

\noindent Really, what evidence do we have which can demonstrate that the Schrödinger equation is exactly solvable within the period of observation for any given system? As a matter of fact, there is plenty of evidence to the contrary. For one, fast and practically realizable exact analytical or numerical solutions to this equation applicable to any given physical system and its associated potential energy are unknown (see, for example,\cite{Ohya} or \cite{Popelier}). As a result, at present the only truly predictive algorithms for solving the Schrödinger equation are ones that built on brute force. Inevitably, the explosion in computational work they lead to warrants only approximate solutions. Moreover, even if there is a superfast algorithm able to exactly solve the Schrödinger equation for any system including that of 500 qubits by manipulating all $2^{500}$ classical bits in a short time $t<T$, any measurement performed after the system has advanced forward by time $T$ will with some probability retrieve a state specified by no more than 500 classical bits, just as in the case of some inexact algorithm, which ignores most of the system's variables and in this way produces an approximate solution to the Schrödinger equation within period $t$.\\

\noindent So, since the statement about vast computational resources of Nature is based on neither a provable proposition nor empirical data but on a mere conjecture \cite{Bernstein}, those what-if questions asked above should be considered at least reasonable to think about.\\

\noindent Obviously, what-if questions will only have answers if one makes some assumptions. This paper makes the assumption that there are no vast computational resources, particularly that solving the Schrödinger equation for any given system is an intractable problem and investigates a role of this assumption in resolving the quantum measurement problem (also known in more recent literature as the problem of the emergence of classicality from quantum systems).\\

\noindent The paper is structured as follows. In section 2 we discuss the computational complexity of the Schrödinger equation and show (using elementary arguments) that solving this equation for any given Hamiltonian is a problem at least as hard as the hardest problems in the \textbf{NP} complexity class, which in turn implies that this problem has no efficient algorithm unless the \textbf{P} complexity class is equal to \textbf{NP}. In section 3 we present two different ways of resolving the quantum measurement problem based on the differing assumptions regarding the \textbf{P }versus \textbf{NP }question -- the approach of the quantum decoherence theory based on the assumption that \textbf{P} = \textbf{NP} and the approach of ours (called computational complexity approach) based on the assumption that\textbf{ P}$\neq$ \textbf{NP}. Section 4 concludes the paper.\\

\section{Quantum computational reductionism}

\noindent To make our discussion on computational complexity more tangible, let us frame the following ``practical'' question:

\begin{quotation}
\noindent \textit{Given a potential} $V\!\!=\!\!V\!\!\left({{\mathbf r}}_1,\dots ,{{\mathbf r}}_N\right)$ \textit{of a system comprised of} $N$ \textit{spin-0 particles of masses} $m_1,\dots ,m_N$, \textit{and a certain level} $E_B$\textit{, is there a state of the system with energy }$E$ \textit{less than or equal to this level }$E_B$\textit{?}
\end{quotation}

\noindent In terms of computational complexity this yes-or-no question is \textit{the} \textit{decision} \textit{problem} (we will refer to it as the problem ${\Pi }_E$) that can be equivalently defined as the set $S_E$ of all inputs for which ${\Pi }_E$ returns 1 (i.e. `yes')

\begin{equation} \label{1}
   S_E=\!\left\{
      \begin{array}{cl}
         \!\!\! N,m_1,\dots ,m_N,V,E_B:
         &
        {\Pi }_E\left(\!
                           \displaystyle -\frac{{\hbar}^2}{2} \sum^N_{n=1} 
                                                            {\frac{{{\nabla }_n}^2}{m_n}}\left.\left|{\psi }_E\!\right.\right\rangle +V\!\!\left.\left|{\psi }_E\!\right.\right\rangle =E\!\!\left.\left|{\psi }_E\!\right.\right\rangle
        \wedge E\le E_B\!\right)\!\!=1 \!\!\!
      \end{array}
   \right\},
\end{equation}
\smallskip

\noindent where ${{\nabla }_n}^2$ denotes the Laplace operator and $\left.\left|{\psi }_E\!\right.\right\rangle $ represents the state of the system with the energy $E$. We will assume that the problem ${\Pi }_E$ is \textit{decidable} (that is, there is an algorithm for ${\Pi }_E$ that instead of looping indefinitely terminates after a finite amount of time and correctly returns a Boolean true 1 or false value 0) and weigh up how difficult the problem ${\Pi }_E$ is to solve.\\

\noindent To begin with, let us notice that the solutions to the problem ${\Pi }_E$ can be quickly verified, namely, one can prove that the state $\left.\left|{\psi }_E\!\right.\right\rangle $ is indeed a solution to ${\Pi }_E$ for a particular instance by substituting $\left.\left|{\psi }_E\right.\right\rangle $ into the expression for ${\Pi }_E$ and then calculating the second derivative values of $\left.\left|{\psi }_E\right.\right\rangle $ (say, with automatic differentiation techniques \cite{Neidinger}) on a deterministic computing device in the polynomial number of steps $T\in O\!\left(N^c\right)$, where $c>0$ is a constant (that does not depend on the particular instance).\\

\noindent Such a property of the problem ${\Pi }_E$ -- verifiability by a deterministic computing device in a polynomial number of steps (i.e. \textit{in polynomial time}) -- suggests that ${\Pi }_E$ belongs to the \textbf{NP} complexity class containing all decision problems for which the instances where the answer is `yes' have efficiently verifiable proofs of the fact that the answer is indeed `yes' (``efficiently verifiable proof'' means a proof by a method each step of which is precisely predetermined and which is certain to produce the answer in a polynomial number of steps).\\

\noindent But can the problem ${\Pi }_E$ be not only efficiently verifiable but also efficiently solvable? Is there a polynomial-time algorithm for ${\Pi }_E$? Before answering to that query, let us put an alternative, more abstract, yes-or-no question:

\begin{quotation}
\noindent \textit{Given a Schrödinger Hamiltonian and an arbitrary condition, is there a solution to the Schrödinger equation subject to this condition?}
\end{quotation}

\noindent It is clear that this problem (we will call it the problem ${\Pi }_P$) is closely related to another one, namely, the \textit{function problem} ${\Phi }_{\psi }$, which is this:

\begin{quotation}
\noindent \textit{Given a Schrödinger Hamiltonian, what is a solution to the Schrödinger equation?}
\end{quotation}

\noindent Accordingly, the problem ${\Pi }_P$ can be defined as the set $S_P$ of inputs (i.e. various Hamiltonians and conditions) for which ${\Pi }_P$ returns 1

\begin{equation} \label{2} 
   S_P=\left\{ 
      \begin{array}{cl}
         H\!\left(t\right),\ P:
         &
         {\rm \ }{\Pi }_P\left(H\!\left(t\right)\!\left.\left|\psi \!\left(t\right)\!\right.\right\rangle =i\hbar \displaystyle  \frac{\partial }{\partial t}\left.\left|\psi \!\left(t\right)\!\right.\right\rangle \ \wedge \ P\left(\left.\left|\psi \!\left(t\right)\!\right.\right\rangle \right)\right)=1
      \end{array}
\right\}\;\;\;\;  , 
\end{equation}
\smallskip

\noindent where $H\!\left(t\right)$ stands for the Hamiltonian (in general, time-dependent) operator and $\!\left.\left|\psi \!\left(t\right)\!\right.\right\rangle$ denotes the Schrödinger equation solution for which the condition $P\left(\left.\left|\psi \!\left(t\right)\!\right.\right\rangle \right)$ holds.\\

\noindent The relationship between the decision problem ${\Pi }_P$ and the function problem ${\Phi }_{\psi }$ is such that if ${\Phi }_{\psi }$ were computable (according to the Church-Turing thesis \cite{Amram}, this means ``if the function $\!\left.\left|\psi \!\left(t\right)\!\right.\right\rangle$ had an algorithm''), then ${\Pi }_P$ would be decidable. Furthermore, if the problem ${\Phi }_{\psi }$ were effectively solvable, then the problem ${\Pi }_P$ would be as well.\\

\noindent Let us assume that the decision problem ${\Pi }_P$ is decidable. Then, to solve an instance of this problem would mean to set up the Hamiltonian for a system accounting for the kinetic and potential energy of the particles constituting the system and having inserted $H\left(t\right)$ into the Schrödinger equation to solve the resulting partial differential (in general, time-dependent) equation for the quantum state $\!\left.\left|\psi \!\left(t\right)\!\right.\right\rangle$ in order to decide whether the ensuing solution $\!\left.\left|\psi \!\left(t\right)\!\right.\right\rangle$ satisfies the condition $P\left(\left.\left|\psi \!\left(t\right)\!\right.\right\rangle \right)$ imposed on the positions, momentums or (and) other physical properties of the system's constituting particles.\\

\noindent Clearly, the problem ${\Pi }_P$ can be quickly modified in the problem ${\Pi }_E$ since the modifying procedure $f:\ S_P\to S_E$, which straightforwardly transforms the problem ${\Pi }_P$ into the problem ${\Pi }_E$

\begin{equation} \label{3}  
   \begin{array}{rcl}
      {f:\ S}_P & \to  & S_E \\ 
      H\left(t\right) & \mapsto  & H= \displaystyle -\frac{{\hbar}^2}{2} \sum^N_{n=1}{\frac{{{\nabla }_n}^2}{m_n}}+V \\ 
      P\left(\left.\left|{\psi }_E\right.\right\rangle \right) & \mapsto  & E\le E_B
   \end{array}
   \;\;\;\; ,
\end{equation}
\smallskip

\noindent can be obviously executed in a polynomial number of steps. This means that the problem ${\Pi }_E$ can be solved using the algorithm for solving ${\Pi }_P$, and for this reason the problem ${\Pi }_E$ is \textit{reducible} to ${\Pi }_P$ (which is intuitively understandable since ${\Pi }_E$ is no more difficult than ${\Pi }_P$). As a consequence, if the problem ${\Pi }_P$ had a polynomial-time algorithm, the problem ${\Pi }_E$ could be efficiently solvable too.\\

\noindent 

\noindent At this point, let us invoke the possibility of encoding a specific instance of a given decision problem in a Hamiltonian. Explicitly, let us consider an adiabatically evolving system characterized by the Hamiltonian $H\!\left(t\right)$, which is slowly varying, and thus at any instant of time $t$ the system remains in the state $\!\left.\left|\psi \!\left(t\right)\!\right.\right\rangle$ close to the instantaneous ground state $\!\left.\left|{\psi }_g\!\left(t\right)\!\right.\right\rangle$ of the Hamiltonian $H\!\left(t\right)$. Suppose we choose $H\!\left(t\right)$ so that at time $t=0\ $the ground state $\!\left.\left|{\psi }_g\!\left(0\right)\!\right.\right\rangle $ of $H\!\left(0\right)$ encodes an input of some decision problem ${\Pi }_C$ ( so $\!\left.\left|{\psi }_g\!\left(0\right)\!\right.\right\rangle $ is known in advance and the system can be easily prepared in $\!\left.\left|{\psi }_g\!\left(0\right)\!\right.\right\rangle $), whereas at time $t=T$ the system's Hamiltonian $H\!\left(T\right)$ coincides with the Hamiltonian $H_C$ whose ground state $\!\left.\left|{\psi }_g\!\left(T\right)\!\right.\right\rangle $ is unknown and encodes the solution to the problem ${\Pi }_C$. If $H\!\left(0\right)$ and $H_C$ are easy to specify, then using the modifying procedure $g:\ S_P\to S_C$

\begin{equation} \label{4}  
   \begin{array}{rcl}
      g:\ S_P & \to  & S_C \\ 
      H\!\left(t\right) & \mapsto  &  \displaystyle \tilde{H}\!\left(t\right)\cong \!\left(1-\frac{t}{T}\right)\!H\!\left(0\right)+\frac{t}{T}H_C \\ 
      P\!\left(\left.\left|\psi \!\left(0\right)\!\right.\right\rangle \right) & \mapsto  & \left.\left|\psi \!\left(0\right)\!\right.\right\rangle =\left.\left|{\psi }_g\!\left(0\right)\!\right.\right\rangle
   \end{array}
\end{equation}
\smallskip

\noindent (where $S_C$ is the set of inputs of ${\Pi }_C$ for which ${\Pi }_C$ returns 1) the algorithm for solving ${\Pi }_P$ can be quickly modified to solve ${\Pi }_C$ (to be precise, to solve the Schrödinger equation $\tilde{H}\!\left(t\right)\!\left.\left|\psi \!\left(t\right)\!\right.\right\rangle ={\mathcal E}\!\left(t\right)\left.\left|\psi \!\left(t\right)\!\right.\right\rangle $ for the system's state $\left.\left|\psi \!\left(t\right)\!\right.\right\rangle $, which at time $t=T$ will be close to the ground state $\left.\left|{\psi }_g\!\left(T\right)\!\right.\right\rangle $ encoding the solution to the given problem ${\Pi }_C$ under the condition that at some initial time $t=0$ the ground state $\left.\left|{\psi }_g\left(0\right)\!\right.\right\rangle $ is known).\\

\noindent As it was demonstrated in the paper \cite{Farhi}, the Hamiltonians $H\!\left(0\right)$ and $H_C$ are straightforward to construct if the decision problem ${\Pi }_C$ is the \textit{Exact Cover}, an\textbf{ NP}-complete problem. Since this particular \textbf{NP}-complete problem can be solved using the algorithm for ${\Pi }_P$, it immediately implies that \textit{every problem} in the \textbf{NP} complexity class can be reducible (in a polynomial number of steps) to ${\Pi }_P$; in other words, it implies that the decision problem ${\Pi }_P$ is \textbf{NP}-complete.\\

\noindent What is more, this infers that the function problem ${\Phi }_{\psi }$ of solving the Schrödinger equation is \textbf{NP}-hard, which means that if we had a polynomial time algorithm (on a deterministic computing device) for finding the solutions to the Schrödinger equation for any given Hamiltonian, we could solve all problems in the \textbf{NP} complexity class in polynomial time.\\

\noindent If such a \textit{quantum computational reductionism} did really take place (i.e. if there were a polynomial algorithm for the \textbf{NP}-hard problem ${\Phi }_{\psi }$), then the complexity class \textbf{P} (which is the set of decision problems solvable on a deterministic computing device within polynomial time) would be equal to the class \textbf{NP}; otherwise, \textbf{P}$\neq$ \textbf{NP}.\\

\noindent In the next section, we will see what role this \textbf{P} versus \textbf{NP} question might play in the resolution of the measurement problem.\\

\section{Two approaches to the measurement problem}

\noindent We now proceed to show one by one two approaches to the measurement problem that are based (tacitly or explicitly) on the opposing assumptions regarding the \textbf{P} versus \textbf{NP} question -- the first is that \textbf{P} = \textbf{NP}, and the next is that \textbf{P}$\neq$\textbf{NP}.\\

\subsection{Quantum decoherence approach}

\noindent Let us consider a Stern-Gerlach experiment, in which an initial spin eigenstate $\left.\left|{\psi }_0\!\right.\right\rangle $ with the eigenvalue $s_x=½$ along the x-axis is separated (by means of a magnetic field that is inhomogeneous along the z-axis) into two orthonormal states $\left.\left|0\right.\right\rangle $ and $\left.\left|1\right.\right\rangle $ that have the respective eigenvalues $s_z=½$ and $s_z=-½$ along the z-axis:

\begin{equation} \label{5}  
   \left.\left|{\psi }_0\!\right.\right\rangle =\frac{1}{\sqrt{2}}\left(\left.\left|0\!\right.\right\rangle +\left.\left|1\!\right.\right\rangle \right)\;\;\;\; ;
\end{equation}
\smallskip

\noindent thus, at the detector one sees either $\left.\left|0\!\right.\right\rangle $ or $\left.\left|1\!\right.\right\rangle $ with the probability of observing the eigenvalue $s_x=½$ again equal to

\begin{equation} \label{6}
   P\left(s_x=½\right)=\frac{1}{2}\;\;\;\; .
\end{equation}
\smallskip

\noindent The quantum decoherence approach to the measurement problem goes as follows: Since there is no consistent formalism to describe the interaction between a quantum and a classical system and since quantum mechanics is a universally applicable theory, every system is basically quantum mechanical. Therefore, to have a consistent theory of measurement, we must treat the detector ${\mathcal A}$ (the measurement apparatus) quantum mechanically. Accordingly, we introduce a Hilbert space ${{\mathcal H}}_{{\mathcal A}{\rm \ }}$ for the detector ${\mathcal A}$ and assume that the orthonormal basis vectors for ${\mathcal A}$ are represented by the exact solutions $\left.\left|\epsilon_k\!\right.\right\rangle $ to the many-body Schrödinger equation with the Hamiltonian $H_{{\mathcal A}}$ describing different configurations $k$ of the detector's $N$ constituent microscopic particles (like different sets of their spatial positions ${{\mathbf r}}_1,{{\mathbf r}}_2,\dots ,{{\mathbf r}}_N$ at the instant of time $t$):

\begin{equation} \label{7}
   i\hbar \frac{\partial }{\partial t}\left.\left|\epsilon_k\!\right.\right\rangle =\left[\sum^N_{n=1}{\frac{{{{\mathbf p}}_n}^2}{2m_n}}+\frac{1}{2M}\sum^N_{j\ne l}{{{\mathbf p}}_j{{\mathbf p}}_l}+{\widehat{{\rm H}}}_{{\rm s}}+V\!\left({{\mathbf r}}_1,{{\mathbf r}}_2,\dots ,{{\mathbf r}}_N,t\right)\right]\!\left.\left|\epsilon_k\!\right.\right\rangle \equiv H_{{\mathcal A}}\!\left.\left|\epsilon_k\!\right.\right\rangle \;\;\;\; , 
\end{equation}
\smallskip

\noindent where the terms of the form ${{\mathbf p}}_j{{\mathbf p}}_l$ (known as mass polarization terms) are due to the kinetic energy dependency on the spatial configuration of the interacting with each other constituent particles ($M$ denotes the mass of the collection of the particles resulting in this extra kinetic energy), ${\widehat{{\rm H}}}_{{\rm s}}$ is the term accounting for the presence of the constituent particles' spins (this terms may include spin-orbit coupling, spin-rotation coupling and spin-spin coupling), and the index $k$ can be continuous, a discrete one, or a mixture of continuous and discrete indexes (in which case the meaning of orthonormality of the basis vectors $\left.\left|\epsilon_k\!\right.\right\rangle $ may turn out to be ambiguous; however, that detail is an inessential for the purposes of this section).\\

\noindent The assumption that the possible states of the macroscopic detector ${\mathcal A}$ are represented by the Hilbert space ${{\mathcal H}}_{{\mathcal A}{\rm \ }}$, whose unit vectors are the exact solutions $\left.\left|\epsilon_k\!\right.\right\rangle $ to the detector's Schrödinger equation, is dictated by the analogy with the Hilbert space ${{\mathcal H}}_{{\mathcal S}{\rm \ }}$ for the observed quantum microscopic system ${\mathcal S}$ -- the spin-$½$ test-particle of mass $m$ and charge $q$ -- spanned by the orthonormal basis vectors $\left.\left|0\!\right.\right\rangle $ and $\left.\left|1\!\right.\right\rangle $

\begin{equation} \label{8}
   \begin{array}{c}
      \left[\!\!
         \begin{array}{c}
            1 \\ 
            0
        \end{array}
      \!\!\right]=\left.\left|0\right.\right\rangle  \\ 
      \left[\!\!
         \begin{array}{c}
             0 \\ 
             1
         \end{array}
      \!\!\right]=\left.\left|1\right.\right\rangle
  \end{array}
  \;\;\;\; , 
\end{equation}
\smallskip

\noindent which are the exact solutions to the Schrödinger equation with the Hamiltonian $H_{{\mathcal S}}$ describing the test-particle flowing through the external inhomogeneous magnetic field ${\mathbf B}$ of the Stern-Gerlach device

\begin{equation} \label{9}
   i\hbar \frac{\partial }{\partial t}
   \left( \!\!\!\!
   \begin{array}{c}
      \left.\left|0\right.\right\rangle  \\ 
      \left.\left|1\right.\right\rangle  
   \end{array}
   \!\!\!\! \right)=
   \left[\left(\frac{{\left({\mathbf p}-q{\mathbf A}\right)}^2}{2m}+q\phi\right){\hat{1}}_{\left[2×2\right]}-\frac{q\hbar }{2m}{\mathbf \sigma }\cdot {\mathbf B}\right]
   \left( \!\!\!\!
   \begin{array}{c}
      \left.\left|0\right.\right\rangle  \\ 
      \left.\left|1\right.\right\rangle  
   \end{array}
   \!\!\!\! \right)
   \equiv H_{{\mathcal S}}
   \left( \!\!\!\!
   \begin{array}{c}
      \left.\left|0\right.\right\rangle  \\ 
      \left.\left|1\right.\right\rangle  
   \end{array}
   \!\!\!\! \right)
   \;\;\;\; , 
\end{equation}
\smallskip

\noindent where the electromagnetic field is defined by the three-component vector potential ${\mathbf A}$ and scalar electric potential $\phi$, while ${\mathbf \sigma }$ is the three-component vector of the Pauli $2×2$ matrices and ${\hat{1}}_{\left[2×2\right]}$ is the $2×2$ identity matrix.\\

\noindent In the conventional treatment, one treats the combined system ${\mathcal S}+{\mathcal A}$ as a closed quantum system (ignoring the environment) with the Hilbert space

\begin{equation} \label{10}
   {\mathcal H}={{\mathcal H}}_{{\mathcal S}{\rm \ }}\!\!\otimes\!\! \ {{\mathcal H}}_{{\mathcal A}{\rm \ }}
   \;\;\;\; . 
\end{equation}
\smallskip

\noindent Writing the total Hamiltonian of the particle-detector combine system ${\mathcal S}+{\mathcal A}$ as

\begin{equation} \label{11}
   H=H_{{\mathcal S}}+H_{{\mathcal A}}+H_{{\rm int}}
   \;\;\;\; , 
\end{equation}
\smallskip

\noindent a standard way to express the interaction term $H_{{\rm int}}$ in the Hamiltonian $H$ is to employ the interaction of the von Neumann form

\begin{equation} \label{12}
   H_{{\rm int}}=
   \left(\!\left.\left|0\!\right.\right\rangle \!\left\langle \left.\!0\right|\right.\!\right)\!\otimes 
   \!\left(\sum_k{A_k\!\left.\left|\epsilon_k\!\right.\right\rangle \!\left\langle \left.\!\epsilon_k\right|\right.}\!\right)
   +
   \left(\!\left.\left|1\!\right.\right\rangle \!\left\langle \left.\!1\right|\right.\!\right)\!\otimes
   \!\left(\sum_k{B_k\!\left.\left|\epsilon_k\!\right.\right\rangle \!\left\langle \left.\!\epsilon_k\right|\right.}\!\right) 
\end{equation}
\smallskip

\noindent (in which $\left.\left|\epsilon_k\!\right.\right\rangle \!\left\langle \left.\!\epsilon_k\right|\right.$ are the operators acting on ${{\mathcal H}}_{{\mathcal A}{\rm \ }}$, $A_k$ and $B_k$ are the definite interaction energies of the $k^{th}$ configuration of the detector's constituent particles for test-particle's eigenstates $\left.\left|0\!\right.\right\rangle $ and $\left.\left|1\right.\right\rangle $ correspondingly) and stipulate that during the interval $[t_i,t_f]$ of the interaction, the interaction term $H_{{\rm int}}$ in the Hamiltonian $H$ dominates over the other two terms, so that, effectively, $H\approx H_{{\rm int}}$.\\

\noindent Let $\left.\left|\epsilon_0\right.\right\rangle $

\begin{equation} \label{13}
   \left.\left|\epsilon_0\right.\right\rangle =\sum_k{a_k\!\left.\left|\epsilon_k\right.\right\rangle }
\end{equation}
\smallskip

\noindent be the initial state of the detector and $U\left(t_f,t_i\right)$

\begin{equation} \label{14}
   U\left(t_f,t_i\right)\ =I-\frac{i\tau }{\hbar }H_{{\rm int}}={\rm exp}\left(-\frac{i\tau }{\hbar }H_{{\rm int}}\right)
\end{equation}
\smallskip

\noindent be the evolution operator for the macroscopic system ${\mathcal S}+{\mathcal A}$ for the short duration $\tau =t_f-t_i$ of the interaction. If at time $t_i$, i.e. before the interaction takes place, the state of the combined system ${\mathcal S}+{\mathcal A}$ is the direct product of the test-particle state and the detector state

\begin{equation} \label{15}
   \left.\left|{\Psi }_i\right.\right\rangle =
   \left.\left|{\psi }_0\right.\right\rangle \!\otimes\! \left.\left|\epsilon_0\right.\right\rangle =
   \frac{1}{\sqrt{2}}\left(\left.\left|0\right.\right\rangle +\left.\left|1\right.\right\rangle \right)\!\otimes\! \sum_k{a_k\!\left.\left|\epsilon_k\!\right.\right\rangle }
   \;\;\;\; , 
\end{equation}
\smallskip

\noindent then linearity of the evolution operator implies that at time $t_f$, i.e. after the interaction has happened, we must get the equation

\begin{equation} \label{16}
   U\left(t_f,t_i\right)\!\left.\left|{\Psi }_i\right.\right\rangle =
   \sum_k{\left(
      \frac{1}{\sqrt{2}}\,a_k\!\left.\left|0\right.\right\rangle \!\otimes\! \left.\left|\epsilon_k\right.\right\rangle e^{-\frac{i\tau }{\hbar }A_k}
      +
      \frac{1}{\sqrt{2}}\,a_k\!\left.\left|1\right.\right\rangle \!\otimes\! \left.\left|\epsilon_k\right.\right\rangle e^{-\frac{i\tau }{\hbar }B_k}
   \!\right)}
   \;\;\;\; , 
\end{equation}
\smallskip

\noindent where the right hand side is a superposition of the quantum states of the macroscopic system ${\mathcal S}+{\mathcal A}$ (in which the detector ``sees'' the test-particle in both $\left.\left|0\right.\right\rangle $ and $\left.\left|1\right.\right\rangle $ states at the same time).\\

\noindent According to the Born probability rule, to compute the probability $P\left(s_x=½\right)$ of observing the spin eigenvalue $s_x=½$ for the test-particle that has made a quantum leap from the initial state $\left.\left|{\psi }_0\right.\right\rangle $ to the final state $\left.\left|{\Psi }_f\right.\right\rangle {\rm =}U\left(t_f,t_i\right)\!\left.\left|{\Psi }_i\right.\right\rangle $, we have to calculate the modulus squared of the scalar product of these two states:

\begin{equation} \label{17}
   P\left(s_x=½\right)=
   {\left|\left\langle {\psi }_0\!\mathrel{\left|\!\vphantom{{\psi }_0 {\Psi }_f}\right.\kern-\nulldelimiterspace}{\Psi }_f\right\rangle \right|}^2=
   \sum_k{{\left|\left\langle \left.{\psi }_0\right|\right.\!\otimes\! \left\langle \left.\epsilon_k\right|\!\left.\left|{\Psi }_f\right.\right\rangle \right.\right|}^2}=
   \sum_k{{\left|a_k\right|}^2{{\cos }^2 \left(\frac{\left(A_k-B_k\right)}{2\hbar }\tau \right)\ }}
   \;\;\;\; . 
\end{equation}
\smallskip

\noindent Now, let us recall that as a macroscopic object, the detector has an enormous ``volume'' available to it in the Hilbert space ${{\mathcal H}}_{{\mathcal A}{\rm \ }}$ corresponding to the detector's microscopic degrees of freedom (which -- due to the interaction between the detector's $N$ internal microscopic particles -- even with the most coarse grain discretization would be of the same magnitude as a double exponential of $N$). This fact might be seen as the practical impossibility of accurately keeping track of the superposition coefficients $a_k$ as well as the interaction energies $A_k$ and $B_k$. So, if initially the detector was in the superposition (\ref{13}) of states such that $a_k\ne 0$ for many values of $k$, then -- \textit{on assumption that the coefficients }$a_k$\textit{ and the interaction energies }$A_k$\textit{ and }$B_k$\textit{ are distributed randomly} -- after a very short period of time the argument of the cosine squared would likely take on several essentially random values. Hence, afterward the weighted average of the cosine squared (with normalized weights $\sum_k{{\left|a_k\right|}^2}=1$) can be replaced by the overall average of sinusoid squared

\begin{equation} \label{18}
   \sum_k{{\left|a_k\right|}^2{{\cos }^2 \left(\frac{\left(A_k-B_k\right)}{2\hbar }\tau \right)\ }}\ {{\underset{\left(\frac{\left(A_k-B_k\right)}{2\hbar }\tau \right)\ \to \ \infty }{\longrightarrow}}}\ \frac{1}{2}
\end{equation}
\smallskip

\noindent giving approximately the classical value to the probability $P\left(s_x=½\right)$

\begin{equation} \label{19}
   P\left(s_x=½\right)=\overline{{{\cos }^2 \left(\frac{\left(A_k-B_k\right)}{2\hbar }\tau \right)\ }}\approx \frac{1}{2}
   \;\;\;\; . 
\end{equation}
\smallskip

\noindent The procedure of averaging the cosine-squared random values over all the possible configurations $k$ (i.e. all the possible sets of the spatial positions) of the detector's microscopic constituent particles can be interpreted as ignoring the detector's microscopic degrees of freedom (which are uncontrolled and unmeasured). Seemingly, this is comparable to the procedure of deriving probability $½$ for `heads' (as well as for `tails') in the experiment of tossing a fair coin by averaging over the uncontrolled and unmeasured degrees of freedom of the environment of the coin \cite{Bub}. However, these two procedures are substantially different. In the coin toss experiment if we take into consideration appropriate environmental parameters, a definite outcome can be predicted. Whereas in the case of the Stern-Gerlach experiment we cannot claim that taking the detector's microscopic degrees of freedom into consideration, a definite outcome of the experiment will be predicted. In fact, doing so we will only get back the non-classical probability (\ref{17}) resulted from the superposition (\ref{16}) of the quantum states of the combined macroscopic system ${\mathcal S}+{\mathcal A}$.\\

\subsection{Computational complexity approach}

\noindent Unlike the equation (\ref{9}) describing the single microscopic test-particle, the Schrödinger equation (\ref{7}) describing $N$ mutually interacting microscopic particles (which constitute a many-body situation) is neither known nor believed to have the exact generic \textit{analytical} solutions (i.e. applicable to an arbitrary many-body system exact solutions constructed using well-known operations that lend themselves readily to calculation of outputs in the short interaction time $\tau $) \cite{Mattis}. Moreover, if the \textbf{NP}-hard problem ${\Phi }_{\psi }$ of finding the solutions to the Schrödinger equation for an arbitrary Hamiltonian was intractable (i.e., if \textbf{P} $\neq$ \textbf{NP}), then the computational effort to solve the Schrödinger equation for an arbitrary system would, in general, scale exponentially with the system's constituent particle number. From whence it would follow that due to the huge number $N$ of the constituent microscopic particles comprising the macroscopic detector, the exact generic \textit{numerical} solutions to the equation (\ref{7}) would be impossible to reach within not only the interaction time $\tau $ but any reasonable amount of time at all (this task would necessarily require vast computational resources that even Nature does not have). Hence, in case of \textbf{P} $\neq$ \textbf{NP} we cannot treat the macroscopic detector quantum mechanically since in that case the orthonormal basis vectors $\left.\left|\epsilon_k\right.\right\rangle $ that span the Hilbert space ${{\mathcal H}}_{{\mathcal A}{\rm \ }}$ for the detector ${\mathcal A}$ cannot be obtainable by any practical means. (This inference may explain why there is a limitation in the application of quantum mechanics to a macroscopic world constituted by small particles obeying the quantum laws.)\\

\noindent On the other hand, an intractable problem can surely be solved in a short time but only if the problem's input is small. This might happen if a system has a small number of the degrees of freedom (like a toy model or a microscopic system completely isolated from the environment) or if the system has just a few effective -- i.e. controlled or measured -- degrees of freedom among many others that are completely ignored (and thus uncontrolled and unmeasured); in the latter case, however, the solution might be only inexact (i.e. with a degree of uncertainty) since the description of the system would be incomplete.\\

\noindent From here, we can infer that to be able to explain the interaction of the microscopic test-particle with the macroscopic detector quantum mechanically, we must ignore the detector's microscopic degrees of freedom in the interacting system ${\mathcal S}+{\mathcal A}$ because only then the Schrödinger equation with the interaction Hamiltonian $H_{{\rm int}}$ would be able to quickly (i.e., in the time not longer than $\tau $) produce the solution $\left.\left|{\psi }_{\tau }\right.\right\rangle $, albeit an inexact one, which would describe (approximately) the final state of the test-particle in practical terms.\\

\noindent An obvious way to do this -- while remaining within the von Neumann measurement scheme -- is to allow significant uncertainties in the interaction energies $A_k$ and $B_k$ associated with the configurations of the detector's microscopic constituent particles such that

\begin{equation} \label{20} 
   \forall k:\ \ \ \ A_k\equiv \tilde{A}+{\alpha }_k\left(\omega \right)\ \ \ ,\ \ \ \ B_k\equiv \tilde{B}+{\beta }_k\left(\omega \right)
   \;\;\;\; , 
\end{equation}
\smallskip

\noindent where $\tilde{A}$ and $\tilde{B}$ are the assigned ``best guess'' estimates for these interaction energies (roughly calculated as proportional to the number of electrons in the detector and inversely proportional to the distance between the test-particle and the detector since the interaction is assumed to be due to the Coulomb force), ${\alpha }_k\left(\omega \right)$ and ${\beta }_k\left(\omega \right)$ are their uncertainties -- the real-valued random (stochastic) functions of equal distribution

\begin{equation} \label{21} 
   \forall k:\ \ \ \ {\alpha }_k\left(\omega \right)\ \sim \ \alpha \left(\omega \right){\rm \ ,\ \ \ \ }\ {\beta }_k\left(\omega \right)\ \sim \ \beta \left(\omega \right) 
\end{equation}
\smallskip

\noindent defined on a set of possible outcomes, the sample space $\Omega $, as

\begin{equation} \label{22} 
   \left\{\ \omega \in \Omega :\ \ \ \left|\alpha \left(\omega \right)\right|\le \tilde{A}{\rm \ ,\ \ \ }\left|\beta \left(\omega \right)\right|\le \tilde{B}\ \right\} 
   \;\;\;\; . 
\end{equation}
\smallskip

\noindent Indeed, permitting such uncertainties in the measurement theory would mean that the interaction energies -- $A_k$ and $A_j$, or $B_k$ and $B_j$ -- identified with the different $\ k\ne j$ sets of spatial positions of the detector's constituent particles would be impossible to differentiate in practical terms. In other words, it would mean that the probability of a certain interaction energy $E_{{\rm int}}$ would be the same for the different configurations $k\ne j$:

\begin{equation} \label{23} 
   \forall E_{{\rm int}},k\ne j:\ \ \ \ \ P\left(A_k\le E_{{\rm int}}\right)=P\left(A_j\le E_{{\rm int}}\right){\rm \ \ ,\ }\ \ P\left(B_k\le E_{{\rm int}}\right)=P\left(B_j\le E_{{\rm int}}\right)
   \;\;\;\; . 
\end{equation}
\smallskip

\noindent It implies that for all practical purposes in the theory of measurement the random variables $A_k$ and $\tilde{A}+\alpha \left(\omega \right)$ would be equal in distribution, i.e., $A_k\ \sim \ \tilde{A}+\alpha \left(\omega \right)$; the same holds for the random variables $B_k$ and $\tilde{B}+\beta \left(\omega \right)$: $B_k\ \sim \ \tilde{B}+\beta \left(\omega \right)$. This gives the following stochastic equalities

\begin{equation} \label{24} 
   \sum_k{A_k\left.\left|\epsilon_k\right.\right\rangle \!\left\langle \left.\epsilon_k\right|\right.}\ \sim \ \left(\tilde{A}+\alpha \left(\omega \right)\right)\hat{1}{\rm \ \ \ ,}\ \ \ 
   \sum_k{B_k\left.\left|\epsilon_k\right.\right\rangle \!\left\langle \left.\epsilon_k\right|\right.}\ \sim \ \left(\tilde{B}+\beta \left(\omega \right)\right)\hat{1}
   \;\;\;\; , 
\end{equation}
\smallskip

\noindent where $\sum_k{\left.\left|\epsilon_k\right.\right\rangle \!\left\langle \left.\epsilon_k\right|\right.=\hat{1}}$ is the identity operator. When substituted into the von Neumann form (\ref{12}), these equalities bring out the stochastic expression for the interaction Hamiltonian $H_{{\rm int}}$ that does not contain the detector's microscopic degrees of freedom

\begin{equation} \label{25} 
   H_{{\rm int}}\!\left(\omega \right)\ \sim \ 
      \left[\left(\tilde{A}+\alpha \!\left(\omega \right)\right)\!\left.\left|0\!\right.\right\rangle \!\left\langle \left.0\right|\right.\!\otimes \!\hat{1}
           +\left(\tilde{B}+\beta   \!\left(\omega \right)\right)\!\left.\left|1\!\right.\right\rangle \!\left\langle \left.1\right|\right.\!\otimes \!\hat{1}\right]
   \;\;\;\; , 
\end{equation}
\smallskip

\noindent the feature that renders the information about the spatial arrangements of the detector's microscopic particles determined by the exact solutions $\left.\left|\epsilon_k\right.\right\rangle $ (unreachable in reasonable time, unless \textbf{P} = \textbf{NP}) immaterial to the inexact (stochastic) solution $\left.\left|{\psi }_{\tau }\!\left(\omega \right)\!\right.\right\rangle $ to the Schrödinger equation with the interaction Hamiltonian $H_{{\rm int}}$

\begin{equation} \label{26} 
   U\!\left(\tau \right)\!\left.\left|{\psi }_0\right.\right\rangle \!\otimes\! \left.\left|\epsilon_0\right.\right\rangle =
   \left(I-\frac{i\tau }{\hbar }\,H_{{\rm int}}\!\left(\omega \right)\right)
      \frac{1}{\sqrt{2}}\left(\left.\left|0\right.\right\rangle +\left.\left|1\right.\right\rangle \right)\!\otimes\! \left.\left|\epsilon_0\right.\right\rangle
   \ \sim 
   \ \left.\left|{\psi }_{\tau }\left(\omega \right)\right.\right\rangle \!\otimes \!\left.\left|\epsilon_0\right.\right\rangle
   \;\;\;\; , 
\end{equation}
\smallskip

\noindent where the solution $\left.\left|{\psi }_{\tau }\left(\omega \right)\right.\right\rangle $ is

\begin{equation} \label{27} 
   \left\{\omega \in \Omega :\ \ \ 
      \left.\left|{\psi }_{\tau }\left(\omega \right)\right.\right\rangle \equiv 
         \frac{1}{\sqrt{2}}\left.\left|0\right.\right\rangle e^{-\frac{i\tau}{\hbar }\left(\tilde{A}+\alpha \left(\omega \right)\right)}
       +\frac{1}{\sqrt{2}}\left.\left|1\right.\right\rangle e^{-\frac{i\tau}{\hbar }\left(\tilde{B}+\beta   \left(\omega \right)\right)} 
   \right\} 
   \;\;\;\; . 
\end{equation}
\smallskip

\noindent At this juncture, let us call to mind that a random variable conceptually does not have a single, fixed value (even if unknown); more exactly, it takes on a set of possible different values (each with an associated probability). Thus, the solution $\left.\left|{\psi }_{\tau }\left(\omega \right)\right.\right\rangle $ obtained in  (\ref{27}) -- which is a complex-valued function of the real-valued random variables $\alpha \left(\omega \right)$ and $\beta \left(\omega \right)$ -- does not represent a single, unique state of the test-particle (or a linear combination, a superposition, of fixed states), rather it represents a set of possible states of the test-particle after its interaction with the detector in a yet-to-be-performed experiment. That is, the solution  (\ref{27}) is the function that associates a possible final state of the test-particle with every instance $\omega $ of the experiment so that $\left.\left|{\psi }_{\tau }\left(\omega \right)\right.\right\rangle $ will vary from instance to instance as the experiment is repeated. Performing the experiment many times, one will find the probability $P\left(s_x=½\right)$ of observing the test-particle's spin eigenvalue $s_x=½$ by calculating in each instance $\omega $ the modulus squared ${\left|\left\langle {\psi }_0\!\mathrel{\left|\vphantom{{\psi }_0 {\psi }_{\tau }\!\left(\omega \right)}\right.\kern-\nulldelimiterspace}{\psi }_{\tau }\left(\omega \right)\right\rangle \right|}^2$ of the scalar product of the initial state $\left.\left|{\psi }_0\right.\right\rangle $ and a possible final state $\left.\left|{\psi }_{\tau }\left(\omega \right)\right.\right\rangle $

\begin{equation} \label{28} 
   \begin{array}{r}
      {\left|\left\langle {\psi }_0\mathrel{\left|\vphantom{{\psi }_0 {\psi }_{\tau }\left(\omega \right)}\right.\kern-\nulldelimiterspace}{\psi }_{\tau }\left(\omega \right)\right\rangle \right|}^2
      \sim
      \left[\displaystyle \frac{1}{2}
                  +
                            \frac{1}{2}{\rm cos}\!\left(\frac{\tilde{A}-\tilde{B}}{\hbar }\tau \right){\rm cos}\!\left(\frac{\alpha \left(\omega \right)-\beta \left(\omega \right)}{\hbar }\tau \right)\right.{\rm } \\
     \left.{\rm -}
          \displaystyle \frac{1}{2}{\sin  \!\left(\frac{\tilde{A}-\tilde{B}}{\hbar }\tau \right)\ }{\sin  \!\left(\frac{\alpha \left(\omega \right)-\beta \left(\omega \right)}{\hbar }\tau \right) }\right]
   \end{array}
\end{equation}
\smallskip

\noindent and afterwards averaging the ensuing real-valued random function ${\left|\left\langle {\psi }_0\!\mathrel{\left|\!\vphantom{{\psi }_0 {\psi }_{\tau }\left(\omega \right)}\right.\kern-\nulldelimiterspace}{\psi }_{\tau }\left(\omega \right)\right\rangle \right|}^2$ over the whole sample space $\Omega $.\\

\noindent To find this average value, let us first find the total span of the random argument of the cosine and sine functions in (\ref{28}) by assessing its min ${\xi }_{{\rm min}}$ and max ${\xi }_{{\rm max}}$ values:

\begin{equation} \label{29} 
   {\xi }_{{\rm max}}=-{\xi }_{{\rm min}}=
      {\mathop{\max }_{\Omega } \left[\frac{\alpha \left(\omega \right)-\beta \left(\omega \right)}{\hbar }\tau \right]\ }
      \approx 
      \frac{\tilde{A}+\tilde{B}}{\hbar }\tau  
   \;\;\;\; . 
\end{equation}
\smallskip

\noindent Assigning probabilities to possible outcomes of this random argument, we choose a uniform probability distribution as there is no reason to favor any one of the propositions regarding the argument's outcomes over the others (in such a case the only reasonable probability distribution would be uniform \cite{Park}, and then the information entropy would be equal to its maximum possible value). Subsequently, the average of the cosine function of the random uniformly distributed argument over $\Omega $ can be estimated as

\begin{equation} \label{30} 
   \overline{{\cos  \left(\frac{\alpha \left(\omega \right)-\beta \left(\omega \right)}{\hbar }\tau \right)\ }}\approx \frac{\hbar }{\left(\tilde{A}+\tilde{B}\right)\tau }\ {\rm sin}\frac{\tilde{A}+\tilde{B}}{\hbar }\tau 
\end{equation}
\smallskip

\noindent (whereas the estimated average of the sine function of the same argument ought to be zero due to the symmetry of the assessed limits ${\xi }_{{\rm min}}$ and ${\xi }_{{\rm max}}$ with respect to 0). Given that the limit ${\mathop{\lim }_{\theta \to \infty } \left({\theta }^{-1}{\sin  \theta \ }\right)\ }$ exists and is equal to 0, one can conclude from (\ref{30}) that

\begin{equation} \label{31} 
   \overline{{\cos  \left(\frac{\alpha \left(\omega \right)-\beta \left(\omega \right)}{\hbar }\tau \right)\ }}{\rm \ }{{\underset{\left(\frac{\tilde{A}+\tilde{B}}{\hbar }\tau \right)\ \to \ \infty }{\longrightarrow}}}\ 0
   \;\;\;\; . 
\end{equation}
\smallskip

\noindent Since the estimated interaction energies $\tilde{A}$ and $\tilde{B}$ for the macroscopic detector are values of classical magnitude, we finally find that after a very short period of time the probability $P\left(s_x=½\right)$ will be of classical form

\begin{equation} \label{32} 
   P\left(s_x=½\right)=\overline{{\left|\left\langle {\psi }_0\mathrel{\left|\vphantom{{\psi }_0 {\psi }_{\tau }\left(\omega \right)}\right.\kern-\nulldelimiterspace}{\psi }_{\tau }\left(\omega \right)\right\rangle \right|}^2}\approx \frac{1}{2} 
   \;\;\;\; . 
\end{equation}
\smallskip

\subsection{Comparing the approaches}

\noindent Comparing (\ref{32}) with the analogous expression (\ref{19}) readily points out where the basic distinction between the approaches lies. In the quantum decoherence theory, the interaction between a particular microscopic system and a related macroscopic system is described at first (i.e. \textit{before} decoherence) deterministically by a way of exactly solving the Schrödinger equation for the interacting systems, and only then (i.e. \textit{after} decoherence) random sampling is brought in to simulate uncertainties (caused by some unknown way, in which the microscopic system is entangled with the macroscopic one) in the combined interacting system. In contrast to this, the computational complexity approach changes such a tactic in the first place solving stochastically the Schrödinger equation (by using a stochastic Hamiltonian, which turns the Schrödinger equation into a stochastic differential equation) that specifies the interaction of the microscopic system with the macroscopic one.\\

\noindent Clearly, such dissimilarity is caused by the different attitude towards the \textbf{P} versus \textbf{NP} question -- particularly, the question of the computational hardness of the Schrödinger equation -- adopted by these two approaches. In fact, the quantum decoherence theory tacitly assumes that the quantum computational reductionism holds -- i.e. that \textbf{P} = \textbf{NP} and thus for any system, including a macroscopic detector, the exact solutions to the Schrödinger equation can be deterministically computed either in an instant or in a time so short (in comparison with the interaction time $\tau $) that it could be ignored in the theory.\\

\noindent By contrast, the computational complexity approach presumes that the problem ${\Phi }_{\psi }$ of solving the Schrödinger equation for any given Hamiltonian is intractable (i.e.\textbf{ P} $\neq$ \textbf{NP}), which implies that to do a parallel with the experiment synchronized calculation (using standard quantum theory) is only possible by stochastically solving the Schrödinger equation for a macroscopic interacting system.\\

\section{Conclusion}

\noindent But then, an objection can be made that in fact, the results of quantum theory, when applied to measurements, by no means depend on the status of the \textbf{P }versus \textbf{NP }question because there are simplified apparatus models, which are solvable (e.g., a von Neumann measurement model of a pointer interacting with a microscopic system) and which give an excellent agreement with experiment. So, the inability to exactly solve the Schrödinger equation for an arbitrary macroscopic system in reasonable time (which might or might not be true) has no consequences for the foundations of the theory.\\

\noindent In order to meet this objection, let us recall that a hard, or intractable, problem is not necessarily a problem for which there is no solution; rather it is a problem for which there are no efficient means of solving. In other words, even though some instances of a hard problem could be guessed (and then verified) in reasonable time, no particular rule is followed how to efficiently solve any other instances of the problem. (For example, even if you have guessed the solution to various instances of the Sudoku puzzle, a \textbf{NP}-complete problem \cite{Yato}, you still will not have an efficient algorithm for solving any new instance of this puzzle.) So, despite the fact that various instances of the Schrödinger equation have been successfully solved, we still do not have an efficient algorithm for solving this equation for an arbitrary Hamiltonian (and hence for an arbitrary system), or any assurance that this equation can be always exactly soluble in reasonable time (such an assurance can be only offered by the proof that \textbf{P }= \textbf{NP}).\\

\noindent In an analogous manner, a number of simple (and because of that) exactly solvable apparatus models (or, for that matter, models of any macroscopic system) cannot guarantee that the mutual orthogonality of the apparatus's state vectors needed to provide the loss of coherence in the modulus squared of the scalar product will always arise in all experiments. But it is quite clear that without such a guarantee the approach of the quantum decoherence theory cannot be stated as a set of the general rules applicable to any physical system. However, to get just such a guarantee for any possible system one should be first required having an efficient algorithm capable of solving the Schrodinger equation exactly for any possible system. Obviously, this could be only achievable if \textbf{P} were to be equal to \textbf{NP}.\\

\noindent Thus, how to choose between the two presented approaches to the problem of the emergence of classicality finally depends on the status of the \textbf{P }versus \textbf{NP }question, a major unsolved problem in computer science.\\

\noindent If, for instance, the equality \textbf{P} = \textbf{NP} were to prove to be correct, then, indeed, Nature would be able to solve the Schrödinger equation for a truly macroscopic system in a moment but soon after the solution would be reached decoherence would make the superposition of the quantum states of the macroscopic system (following from the linearity of the Schrödinger equation) unavailable for inspection by local observers. However, if \textbf{P} were to turn out to be not equal to \textbf{NP}, then there would be no physical means to solve this equation for the macroscopic system within a reasonable amount of time; hence, the superposition of the macroscopic system's quantum states as a linear combination of the exact solutions to the Schrödinger equation for this system would be originally nonexistent and thus unavailable for inspection by any observer.\\

\end{document}